\documentclass[conference]{IEEEtran}
\usepackage{cite}
\usepackage{amsmath,amssymb,amsfonts}
\usepackage{algorithmic}
\usepackage{graphicx}
\usepackage{textcomp}
\usepackage[table]{xcolor}
\usepackage{multirow}
\usepackage{pifont}
\usepackage{array}
\usepackage{caption}
\usepackage{subcaption}
\usepackage{mathabx}
\usepackage{array}
\usepackage{mdwmath}
\usepackage{mdwtab}
\usepackage{eqparbox}
\usepackage{url}
\usepackage{authblk}
\usepackage{fancyhdr} 
\usepackage[ruled,vlined]{algorithm2e}
\usepackage[utf8]{inputenc}
\usepackage{helvet} 

\DeclareGraphicsExtensions{.pdf,.png,.jpg}

\usepackage{fancyhdr} 
\fancypagestyle{plain}{%
	\fancyhf{} 
	\fancyfoot[C]{\thepage} 
}

\newcommand{\cmark}{\ding{51}}
\newcommand{\xmark}{\ding{55}}

\def\BibTeX{{\rm B\kern-.05em{\sc i\kern-.025em b}\kern-.08em
		T\kern-.1667em\lower.7ex\hbox{E}\kern-.125emX}}

\fancypagestyle{firstpage}{%
 \fancyhf{}%
 \fancyhead[L]{CPSAT 2024 - The 5th CPSSI International Symposium on Cyber-Physical Systems (Applications and Theory)}
  
 \fancyfoot[L]{%
  \normalsize{XXX-X-XXXX-XXXX-X/XX/\$XX.00 ~\copyright2024 IEEE}
 }%
}
\begin{document}
	
	\title{EdgeLinker: Practical Blockchain-based Framework for Healthcare Fog Applications to Enhance Security in Edge-IoT Data Communications\\
	}

	 \author[$\star$]{Mahdi Akbari Zarkesh}
	 \author[$\star$]{Ehsan Dastani}
	 \author[$\star$]{Bardia Safaei}
	 \author[$\star$]{Ali Movaghar\vspace{-10pt}}
	 \affil[$\star$]{Department of Computer Engineering, Sharif University of Technology, Tehran, Iran}
	 \affil[ ]{Email: \{mahdi.akbarizarkesh,  ehsan.dastani98, bardiasafaei, movaghar\}@sharif.edu\vspace{-10pt}}

	 \maketitle
	\thispagestyle{firstpage}

	\begin{abstract}
		The pervasive adoption of Internet of Things (IoT) has significantly advanced healthcare digitization and modernization. Nevertheless, the sensitive nature of medical data presents security and privacy challenges. On the other hand, resource constraints of IoT devices often necessitates cloud services for data handling, introducing single points of failure, processing delays, and security vulnerabilities. Meanwhile, the blockchain technology offers potential solutions for enhancing security, decentralization, and data ownership. An ideal solution should ensure confidentiality, access control, and data integrity while being scalable, cost-effective, and integrable with the existing systems. However, current blockchain-based studies only address \textbf{som}e of these requirements. Accordingly, this paper proposes EdgeLinker; a comprehensive solution incorporating Proof-of-Authority consensus, integrating smart contracts on the Ethereum blockchain for access control, and advanced cryptographic algorithms for secure data communication between IoT edge devices and the fog layer in healthcare fog applications. This novel framework has been implemented in a real-world fog testbed, using COTS fog devices. Based on a comprehensive set of evaluations, EdgeLinker demonstrates significant improvements in security and privacy with reasonable costs, making it an affordable and practical system for healthcare fog applications. Compared with the state-of-the-art, without significant changes in the write-time to the blockchain, EdgeLinker achieves a 35\% improvement in data read time. Additionally, it is able to provide better throughput in both reading and writing transactions compared to the existing studies. EdgeLinker has been also examined in terms of energy, resource consumption and channel latency in both secure and non-secure modes, which has shown remarkable improvements.
	\end{abstract}
	
	
	\section{Introduction} \label{section.Introduction}
		\noindent
		Historically, medical data has always been of special importance in terms of processing and security. Today, due to the rapid growth of the Internet of Things (IoT) and the advancement of data processing, this issue has become even more significant. This is because, nowadays, doctors use IoT devices capable of collecting medical data to monitor the health status of patients. IoT is a communication infrastructure that connects numerous resource-constrained embedded devices with unique identifiers to each other through internet-based communication technologies, without human intervention \cite{b0,b1,b20}. IoT devices have specific identifiers, enabling them to receive various information from the environment, network, and generally the internet, and perform necessary processing on it. Furthermore, they can connect to a cloud service provider if they lack sufficient processing resources \cite{b21,b22}.
		
		Due to the rapid growth of IoT devices and the need for real-time processing, cloud services are often unsuitable because of high latency \cite{b23}. It is essential that processing occurs as close as possible to the end devices. This fundamental requirement has led to the concept of fog computing, which signifies a paradigm shift in data processing and analysis methods at the network edge \cite{b2}. By extending cloud capabilities to the network edge, fog computing enables real-time processing, reduces data transmission latency, and enhances system reliability. Therefore, the fog layer adds a new dimension to the traditional architecture of cloud and IoT devices.

		Despite the advantages that fog computing offers, there remain risks of unauthorized access and data modification for the data processed and stored at this layer. For example, many studies have discussed how data is processed in the fog layer, but they have not addressed potential vulnerabilities that could lead to intentional or unintentional data modification, which impacts the processing results. One of the solutions that can tackle this issue is the use of blockchain technology and smart contracts. This technology allows us to maintain a shared database without the need for a central authority, thereby providing transparency. Transparency ensures accountability and reduces the risk of fraud and data manipulation. Furthermore, blockchain provides traceability, enabling users to track the origin and movement of assets throughout their lifecycle. The decentralized nature of blockchain makes it highly resilient against data integrity breaches and certain vulnerabilities \cite{b3}. Blockchain, instead of relying on a centralized server susceptible to a single point of failure, distributes data across multiple nodes in the network. This allows nodes that do not trust each other to interact securely \cite{b4}.
		
		One of the major challenges in employing blockchain and smart contracts in IoT applications is how to integrate them in a scalable and efficient manner so that we can effectively utilize the capabilities of blockchain in IoT healthcare applications. Consequently, the applicability of the proposed solutions must be examined via real-world implementations so that their accuracy can be thoroughly examined. Therefore, this paper proposes EdgeLinker; a framework, that demonstrates the applicability of the proposed solutions in the healthcare domain and their potential use for patient health monitoring. In this framework, IoT devices send the patient's information, e.g., the average heart rate, to the network every minute. On the other hand, the doctor intends to retrieve the patient's information history to make better-informed decisions. For this purpose, an access-controlled smart contract is used, which stores the user's heart rate information in an array and grants access only to authorized individuals. First, a smart contract is created on the Ethereum blockchain for data storage. The patient then sends data through a secure communication channel to the miners. If the information is verified and the user has the necessary access, the data is stored in the smart contract. The user can request to grant access to the data to an individual or organization at any time. Additionally, the user can revoke access at any time. When individuals request to read the data, they send a transaction through the secure communication channel to the miners, and if they have the necessary access rights, they can retrieve the data. 
		
		Due to the nature of blockchain, it is important to minimize the read and write times as much as possible to enhance the scalability of our proposed system. So, according to an extensive set of real-world experiments, EdgeLinker has shown to be efficient in terms of many aspects. For example, it has achieved 35\% improvement in read time compared to state-of-the-art, with nearly zero difference in data write time on the blockchain. Additionally, EdgeLinker demonstrates better scalability due to its linearly improved throughput with the increase in the number of fog nodes. This has been achieved by integrating a Proof-of-Authority consensus mechanism on the Ethereum blockchain, which enables access control through smart contracts. Additionally, advanced cryptographic algorithms are employed to secure data communication between IoT edge devices and the fog layer in healthcare applications. The framework's deployment is managed using Docker Swarm, which ensures efficient load balancing and orchestration of the fog nodes. Furthermore, we observed a 1.32ms time overhead for message transmission in the secure channel, which is only 0.2ms more than the non-secure channel. this increase in time is only about 0.2ms, which is negligible compared to the total processing time. As a result, we can use a secure channel to enhance the security of our framework without sacrificing scalability. This framework has been shown to achieve lower CPU and RAM consumption compared to other works.
		
		The rest of this paper is organized as follows: in Section \ref{section.related}, the major studies will be reviewed and evaluated from different perspectives. Then, in Section \ref{section.methodology}, the structure of the proposed framework will be explained in detail. In Section \ref{section.eval}, the results obtained from this design will be examined. Finaly, the paper will be concluded in Section \ref{section.conclusion}.

	\begin{table*}[t]
		\caption{{\small Summary of Related Research.}}
		\label{table1}
		\centering
		\renewcommand{\familydefault}{\sfdefault} 
		\fontfamily{phv}\selectfont 
		\resizebox{\textwidth}{!}{%
		\begin{tabular}{|c|c|c|c|c|c|c|c|c|c|c|}
			\hline
			
			\multirow{2}{*}{\textbf{Author(s)}} & \multicolumn{3}{|c|}{\textbf{Integration}} & \multirow{2}{*}{\textbf{Platform-Independent}} & \multicolumn{3}{|c|}{\textbf{Security Features}} & \multirow{2}{*}{\textbf{Versatility}} & \multirow{2}{*}{\textbf{Operational Feasibility}} & \multirow{2}{*}{\textbf{Decentralized Management}} \\
			
			\cline{2-4}
			\cline{6-8}
			
			 & \textbf{IoT} & \textbf{Fog} & \textbf{Cloud} &  & \textbf{Integrity} & \textbf{Authentication} & \textbf{Confidentiality} &  &  &  \\
			\hline
			
			Chen et al.\cite{b12} & \cmark & \cmark & \cmark & \xmark & \xmark & \xmark & \xmark & \xmark & \xmark & \xmark \\
			\hline
			
			Bruneo et al.\cite{b5} & \cmark & \cmark & \xmark & \cmark & \xmark & \cmark & \cmark & \cmark & \cmark & \xmark \\
			\hline
			
			Yi et al.\cite{b13} & \cmark & \cmark & \cmark & \cmark & \xmark & \cmark & \xmark & \cmark & \cmark & \cmark \\
			\hline
			
			Liang et al.\cite{b6} & \cmark & \xmark & \cmark & \xmark & \xmark & \cmark & \cmark & \xmark & \xmark & \xmark \\
			\hline
			
			Shen et al.\cite{b14} & \cmark & \cmark & \xmark & \cmark & \cmark & \cmark & \cmark & \xmark & \xmark & \cmark \\
			\hline
			
			Vora et al.\cite{b9} & \cmark & \cmark & \cmark & \xmark & \xmark & \cmark & \cmark & \xmark & \xmark & \cmark \\
			\hline
			
			Azaria et al.\cite{b15} & \cmark & \xmark & \cmark & \xmark &\cmark & \cmark & \cmark & \xmark & \xmark & \cmark \\
			\hline
			
			Bhattacharya et al.\cite{b16} & \cmark & \cmark & \cmark & \cmark & \cmark & \cmark & \cmark & \xmark & \cmark & \cmark \\
			\hline
			
			Yazdinejad et al.\cite{b17} & \cmark & \xmark & \cmark & \xmark & \cmark & \xmark & \xmark & \xmark & \xmark & \cmark \\
			\hline
			
			Sharma et al.\cite{b18} & \cmark & \xmark & \cmark & \xmark & \cmark & \cmark & \cmark & \xmark & \cmark & \cmark \\
			\hline
			
			Ouyang et al.\cite{b19} & \cmark & \cmark & \cmark & \xmark & \cmark & \cmark & \xmark & \xmark & \cmark & \cmark \\
			\hline
			
			EdgeLinker & \cmark & \cmark & \cmark & \cmark & \cmark & \cmark & \cmark & \cmark & \cmark & \cmark \\
			\hline
			
		\end{tabular}}
		\fontfamily{\rmdefault}\selectfont 
      \vspace{-10pt}

	\end{table*}
	
	\section{Related Works} \label{section.related}
		\noindent
		In the healthcare industry, protecting data security, managing transactions, and ensuring the indivisible integrity of data are essential components for any data-driven organization. Blockchain technology has the potential to effectively address these critical concerns. Bruneo et al \cite{b5} presented a fog-focused framework named Stack4Things as a PaaS for deploying and executing multiple applications related to IoT devices with optimal computing. This program enables users to manage IoT infrastructure, remotely control nodes, virtualize their functions, and establish network overlays between them.
		
		Liang et al\cite{b6} focused on cloud services in their research. To effectively identify user behaviors and collect data from the source, they used cloud records as the data unit. Their study resulted in the creation of ProvChain. To collect and validate the provenance of cloud data, this architecture embeds provenance data into blockchain transactions. The three stages of the ProvChain architecture are collecting data from the source, storing data from the source, and validating data from the source.
		
		The healthcare industry is continuously evolving with new blockchain-based innovations that encompass multiple data sources, blockchain technology, healthcare applications, and stakeholders. Gordon and Catalini's review \cite{b7} discusses how these layers are organized at different levels. Their research indicates that blockchain technology can revolutionize the healthcare industry by shifting control from institutions to patients.
		
		Sial et al \cite{b8} claim that blockchain technology and smart contracts, which can increase the speed of the entire process, can be beneficial for the healthcare industry. The authors believe that by securely storing data in a ledger, blockchain can reduce the likelihood of data loss and prevent information tampering.
		
		Research conducted by Vora et al \cite{b9} addresses the issue of compromising patient information, including private details such as names and addresses. To overcome this problem, the researchers proposed a blockchain-based system for controlling electronic health records. Their primary goal was to evaluate how well their proposed framework meets the needs of patients, doctors, and other relevant parties.
		
		Zhang et al, in a paper \cite{b10}, explored the potential of blockchain and smart contracts in addressing healthcare challenges. They examined the early stages of integrating blockchain technology in several healthcare use cases and highlighted deployment issues. Their study demonstrates the potential of blockchain-based technologies for more successful healthcare management.
		
		A review of previous research shows that blockchain and similar technologies have been widely studied for their potential to enhance data management, privacy, and security in the healthcare sector. Data breaches, unauthorized access, and data integrity issues may be addressed thanks to the decentralization, transparency, and immutability of blockchain. Table \ref{table1} provides a summary of previous research, examining various parameters in each study.

	\section{Proposed Methodology} \label{section.methodology}
		\noindent
		To achieve the stated goals of ensuring data integrity and system authentication, along with integrating IoT data with blockchain, we aim to introduce a new framework called EdgeLinker in this section. We will describe the various components of the proposed framework and explain the necessity of each part.
		
		\subsection{Architecture} \label{section.methodology.architecture}
			\begin{figure}
				\includegraphics[width=\linewidth]{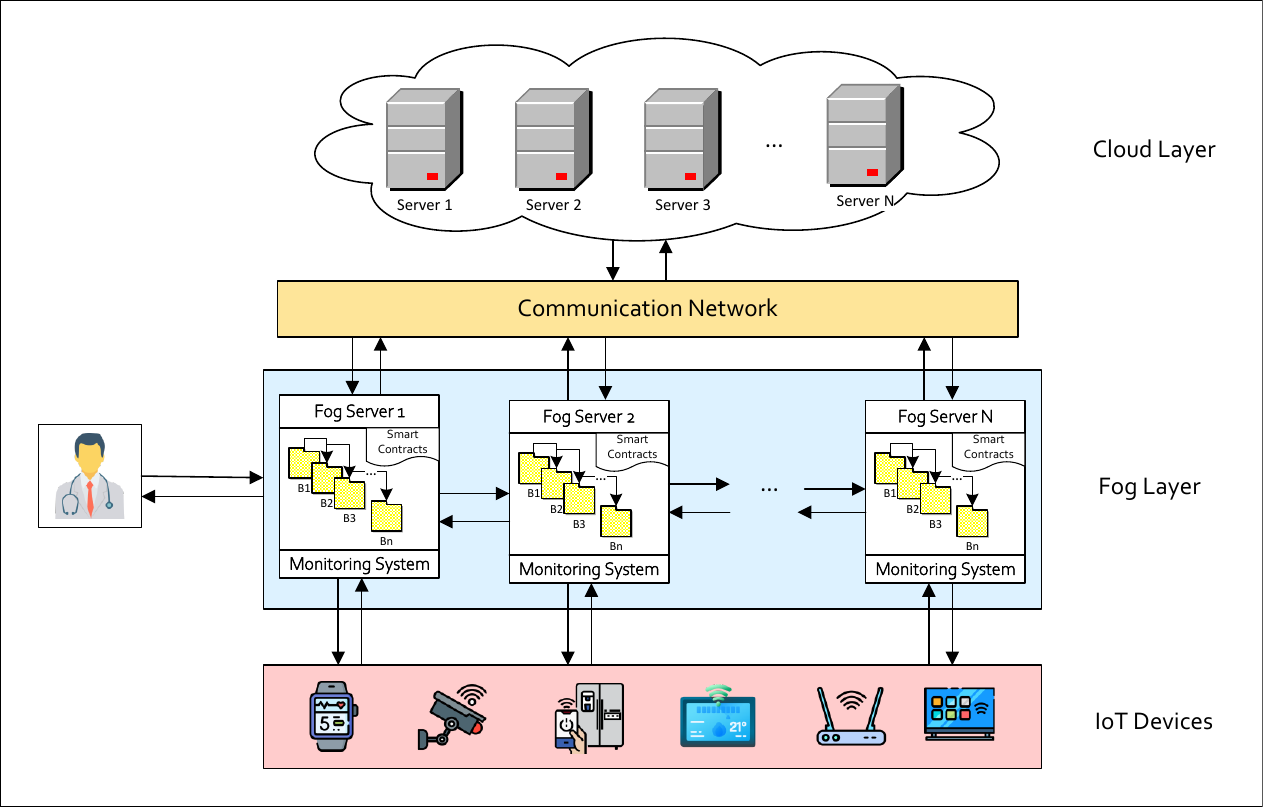}
				\caption{{\small Architecture of EdgeLinker: Cloud Data Center, Fog Layers, IoT Devices.}}
				\label{figure1}
    \vspace{-10pt}
			\end{figure}
		
		\noindent
		To leverage the computational capacity of fog computing at the network edge and reduce service latency, a three-layer architecture is required. Figure \ref{figure1} shows the three-layer architecture of the proposed framework, which includes a cloud data center, fog layers, and IoT devices. The cloud data center can collect data from the fog layer if needed, perform complex computations on it, and send the results back to the fog layer. The next layer consists of fog computing devices, which include the blockchain ledger, smart contracts, and a comprehensive monitoring system. Fog nodes act as miners to collect, process, and validate blocks and messages, creating a distributed computing environment. They also verify new blocks to prevent fraudulent transactions and maintain blockchain integrity. Finally, in the last layer, IoT devices are equipped with sensors, actuators, and other data collection techniques to gather essential and critical data from the physical world. The generated data is then sent to the fog layer for further processing. 
		
		Now that we are familiar with the general architecture of the proposed framework, we will examine each component in more detail and then discuss the interactions between the different components more precisely.
			
		\subsection{Fog Layer} \label{section.methodology.foglayer}
			\noindent
			In the fog layer, we have the blockchain ledger, smart contracts, and a comprehensive monitoring system. Each fog node acts as a miner and, upon receiving a message, first validates it and then creates a new block using the received messages. The new block is then sent to other fog nodes for validation.
			
			\subsubsection{Blockchain} \label{section.methodology.foglayer.bc}
				\noindent
				Given the high importance and sensitivity of medical data and the need to maintain the integrity and accuracy of this data, we intend to use the blockchain data structure as a means for distributed storage of this information. Additionally, to automate existing processes and enable the use of the proposed framework in a personalized manner, our blockchain must support user-customized applications. Therefore, the proposed framework should support smart contracts. Finally, the proposed framework must be characterized by good speed and scalability. Considering these requirements, we plan to use the Proof of Authority (PoA) consensus algorithm in our blockchain. This algorithm is an effective method that allows only authorized nodes, also known as authority nodes, to participate in the block mining process.
				
				To prevent unauthorized access and reading from the blockchain, a semi-private network is used to implement the proposed framework. In this network, only miners can view the network's history for validation purposes, and the history is not visible to others. It is worth noting that this does not interfere with the operation of IoT devices, which can connect to the network whenever needed, call the required smart contracts, and then disconnect from the network. Additionally, events related to smart contracts, the creation of new blocks, and notifications of changes can be tracked from outside the fog layer.
				
				The structure of the desired blocks includes a timestamp of block creation, user messages containing signed hashes, and a hash of the previous block. When miners intend to create a new block, they collate the received user messages. They then place the hash of the previous block in the header of the new block and generate a hash of the entire block, which they sign with their private key. Finally, the new block is broadcasted to the entire network for validation, and if approved, the block is added to the ledger of the fog nodes.
				
				Maintaining and executing the blockchain by miner nodes requires an incentive system so that individuals provide their computational resources for execution and processing. Payment using a local coin is a fair solution to ensure that IoT devices have equitable access to computational resources, while miners are incentivized to validate transactions and support network operations, thereby increasing network stability.
				
			\subsubsection{Smart Contracts} \label{section.methodology.foglayer.sc}
				\noindent
				Smart contracts play a vital role in automating processes; however, it should be noted that to maintain data confidentiality, all smart contracts within the proposed framework that intend to store data must utilize the access control system that will be introduced later.
				
				\begin{algorithm}[t]
					\caption{Permission-Based Access Control}
					\label{algorithm1}
					\SetKwFunction{FInitialize}{\textsc{Initialize}}
					\SetKwFunction{FHasPermission}{\textsc{HasPermission}}
					\SetKwFunction{FGrantPermission}{\textsc{GrantPermission}}
					\SetKwFunction{FRevokePermission}{\textsc{RevokePermission}}
				
					\BlankLine
					
					\SetKwProg{Fn}{Function}{:}{}
					\SetKwProg{Proc}{Procedure}{:}{}
					\SetKw{KwTo}{in}
					\SetKw{KwRet}{return}
					
					Declare \textit{\_permission} as a mapping from Permission to List[Address]\;
					Declare PERMITTER\_PERMISSION as $0x00$\;
					\BlankLine
					
					\textbf{Input: }transaction T
					
					\Fn{\FInitialize{}}{
						\textit{\_permission}[PERMITTER\_PERMISSION].add(T.sender)\;
					}
					\BlankLine
					
					\Fn{\FHasPermission{permission, address}}{
						\KwRet \textit{\_permission}[\textit{permission}].has(address)\;
					}
					\BlankLine
					
					\Proc{\FGrantPermission{permission, address}}{
						\If{\FHasPermission{PERMITTER\_PERMISSION, T.sender}}{
							\textit{\_permission}[\textit{permission}].add(address)\;
						}
					}
					\BlankLine
					
					\Proc{\FRevokePermission{permission, address}}{
						\If{\FHasPermission{PERMITTER\_PERMISSION, T.sender}}{
							\textit{\_permission}[\textit{permission}].remove(address)\;
						}
					}
				\end{algorithm}

				Algorithm \ref{algorithm1} illustrates access control in smart contracts using access permissions. This method restricts access to specific data and functions to authorized individuals only, thereby enhancing the confidentiality of the stored data.
				
				To automate existing processes in traditional systems, it is possible to use ``events" that can be tracked by applications running on cloud servers and other network nodes, enabling indirect interaction. By utilizing events, it is feasible to implement pipelines from the data collection stage to the time of processing and delivery. Implementing these pipelines provides a secure and reliable approach for collecting data from sensors and IoT devices, such as heart rate monitors, medical devices, and other types of data sensors, and then processing the data. Additionally, with the help of oracles, it is possible to receive data from the external environment and make decisions based on that data.
				
				\subsubsection{Comprehensive Monitoring System} \label{section.methodology.foglayer.monitoring}
					\noindent
					One of the features of smart contracts is that to activate and execute them, a message must be sent to their address, which will automatically activate if the conditions are met. Additionally, these contracts only have access to the data sent to them or predefined data. This can be problematic in cases such as detecting fraudulent patterns and security issues, where the system needs to examine all incoming and outgoing transactions and send alerts if necessary. For this purpose, the present study utilizes a comprehensive monitoring system.
					
					This monitoring system is installed and deployed on all existing fog nodes. In the initial version, this system only includes sending an alert if an invalid new block is detected across the entire network. However, in future versions, we intend to add more complex and precise conditions for detecting invalid behaviors.

			\subsection{Secure Communication Channel} \label{section.methodology.channel}
				\noindent
				In this research, a communication channel is designed to ensure the identification, authentication, and authorization of devices, as well as the proper transmission of messages. The proposed solution should be cost-effective and minimize the volume of transmitted messages. Given that asymmetric encryption of entire messages increases their size, and symmetric encryption algorithms benefit from hardware implementation, thereby improving speed, we intend to use a combination of these two methods in this research. The proposed solution should also be able to prove the sender's message transmission. Therefore, all entities will have a pair of public and private keys. The public key is used to identify each entity's identity and is publicly accessible. However, the private key is confidential and only held by the entity itself, used for decrypting data. All messages sent in the system include a timestamp, nonce, and identification field. Each public key in the network has a nonce value that increments by one with each transaction sent. This helps prevent replay attacks and maintains the order of transactions. The identification value is the individual's public key, used to identify the individuals.

				\begin{algorithm}[t]
					\caption{Message Transmission Through Secure Channel}
					\label{algorithm2}
					\KwIn{
						\begin{itemize}
							\item $m$: Message with Nonce to be sent
							\item $SK_s$: Sender's private key
							\item $PK_r$: Receiver's public key
						\end{itemize}
					}
					\KwOut{
						\begin{itemize}
							\item $c$: Encrypted and signed message
						\end{itemize}
					}
					\BlankLine
					
					Compute shared key $K = DH(SK_s, PK_r)$ using authenticated Diffie-Hellman\;
					Compute hash $H(m)$ of message\;
					Sign hash: $sig \gets Sign(H(m), SK_s)$\;
					Concatenate message and signature: $m' \gets m \parallel sig$\;
					Encrypt concatenation: $c \gets Encrypt(m', K)$\;
					Send encrypted message $c$ to receiver\;
					
				\end{algorithm}
					
				\begin{algorithm}[t]
					\caption{Message Reception Through Secure Channel}
					\label{algorithm3}
					\KwIn{
						\begin{itemize}
							\item $c$: Encrypted and signed message
							\item $K$: Shared DH key
							\item $PK_s$: Sender's public key
						\end{itemize}
					}
					\KwOut{
						\begin{itemize}
							\item $m$: Original message if authentication succeeds
							\item Error otherwise
						\end{itemize}
					}
					\BlankLine
					
					Decrypt with shared key: $c' \gets Decrypt(c, K)$\;
					Parse $c'$ as $m \parallel sig$\;
					Compute $h \gets Hash(m)$\;
					Compute sender hash: $h' \gets Decrypt(sig, PK_s)$\;
					\If{$h' = h$}{
						\Return $m$\;
					}
					\Else{
						\Return Error\;
					}
					
				\end{algorithm}

				Before sending a message, the sender performs the steps outlined in Algorithm \ref{algorithm2}, which include: computing the message hash, signing the computed hash with the private key for authentication and appending it to the end of the message, encrypting the entire message using the Diffie-Hellman key to prevent eavesdropping on the message content, and then sending the message to the recipient.
				
				On the receiver's side, the steps outlined in Algorithm \ref{algorithm3} are followed: first, the message is decrypted using the Diffie-Hellman key, and then the hash value of the received message is computed. Next, the received hash along with the message is decrypted using the sender's public key. If the hash values from both steps match, the received message has been correctly delivered to the recipient, and the sender is authenticated. After receiving the message, the received nonce value is processed with the last stored value in the blockchain to check for repetition and to maintain the order of messages.
				
				To support legacy devices, these devices can utilize the computational resources of the fog layer as an intermediary or proxy to send their requests to the network. This method ensures the security of data messages in legacy devices to some extent and allows for the utilization of existing device capabilities.

	\section{Evaluation and Analysis} \label{section.eval}
		\noindent
		In this section, we will discuss the system setup used to evaluate the performance of the proposed system. In this design, the front-end was implemented using React Native, a Web3-compatible framework that provides direct communication with the blockchain. For the back-end, we used Node.js 18.16, and the contracts were implemented in the Remix IDE + Thor environment using Solidity. In this design, the Sync 2 wallet was chosen for its support of various networks and easy-to-use interface. Additionally, to implement the proposed framework, we used a modified version of the IBFT 2.0 algorithm, as this algorithm is resilient to node failures.
		
\begin{figure}
			\includegraphics[width=\linewidth]{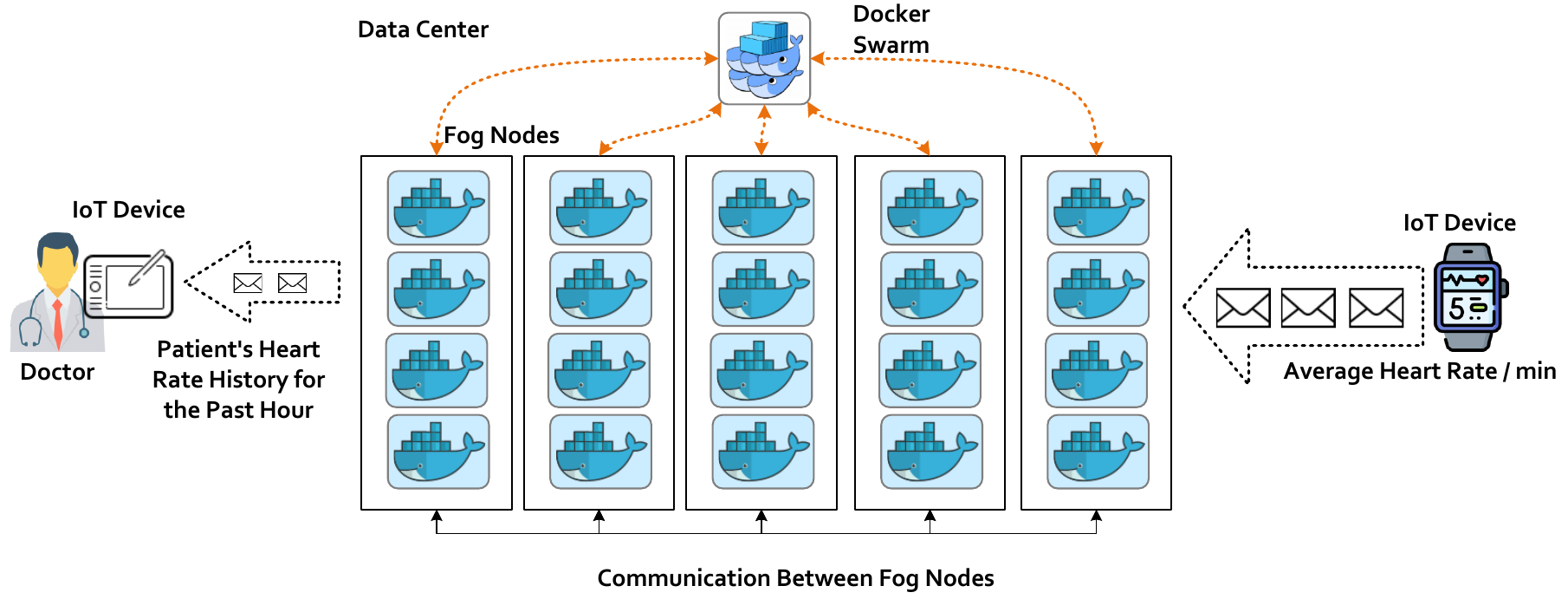}
			\caption{{\small Experimental Environment.}}
			\label{finalfigure}
      \vspace{-13pt}
		\end{figure}

		To evaluate the performance of the proposed architecture, experiments were conducted using hardware specifications that include 5 cloud servers in a data center, each with 16 CPU cores and 64GB of RAM. Docker Swarm was used for load management, and each fog node was allocated 2 CPU cores and 8GB of RAM. Figure \ref{finalfigure} shows a view of the described experimental environment. Moreover, Table \ref{table3} provides a summary of the system setup required for this architecture.
		
		\begin{table}[t]
			\caption{{\small System Setup.}}
			\label{table3}
			\centering
            \renewcommand{\familydefault}{\sfdefault} 
			\fontfamily{phv}\selectfont 
			\begin{tabular}{|l|l|}
				\hline
				\textbf{front-end} & React Native\\
				\hline
				\textbf{back-end} & Node.js 18.16\\
				\hline
				\textbf{Blockchain} & Ethereum\\
				\hline
				\textbf{Wallet} & Sync 2\\
				\hline
				\textbf{Deployment of Smart Contracts} & Remix IDE + Thor\\
				\hline
				\textbf{Consensus Algorithm} & IBFT 2.0\\
				\hline
				\textbf{Fog Node CPU Cores} & 2\\
				\hline
				\textbf{Fog Node RAM} & 8GB\\
				\hline
				\textbf{IoT device} & Galaxy Watch 4 Classic\\
				\hline
				\textbf{IoT device} & Samsung Tablet\\
				\hline
				\textbf{Communication Technology} & Wi-Fi\\
				\hline
			\end{tabular}
            \fontfamily{\rmdefault}\selectfont 
		\end{table}
		
		\begin{figure}
			\includegraphics[width=\linewidth]{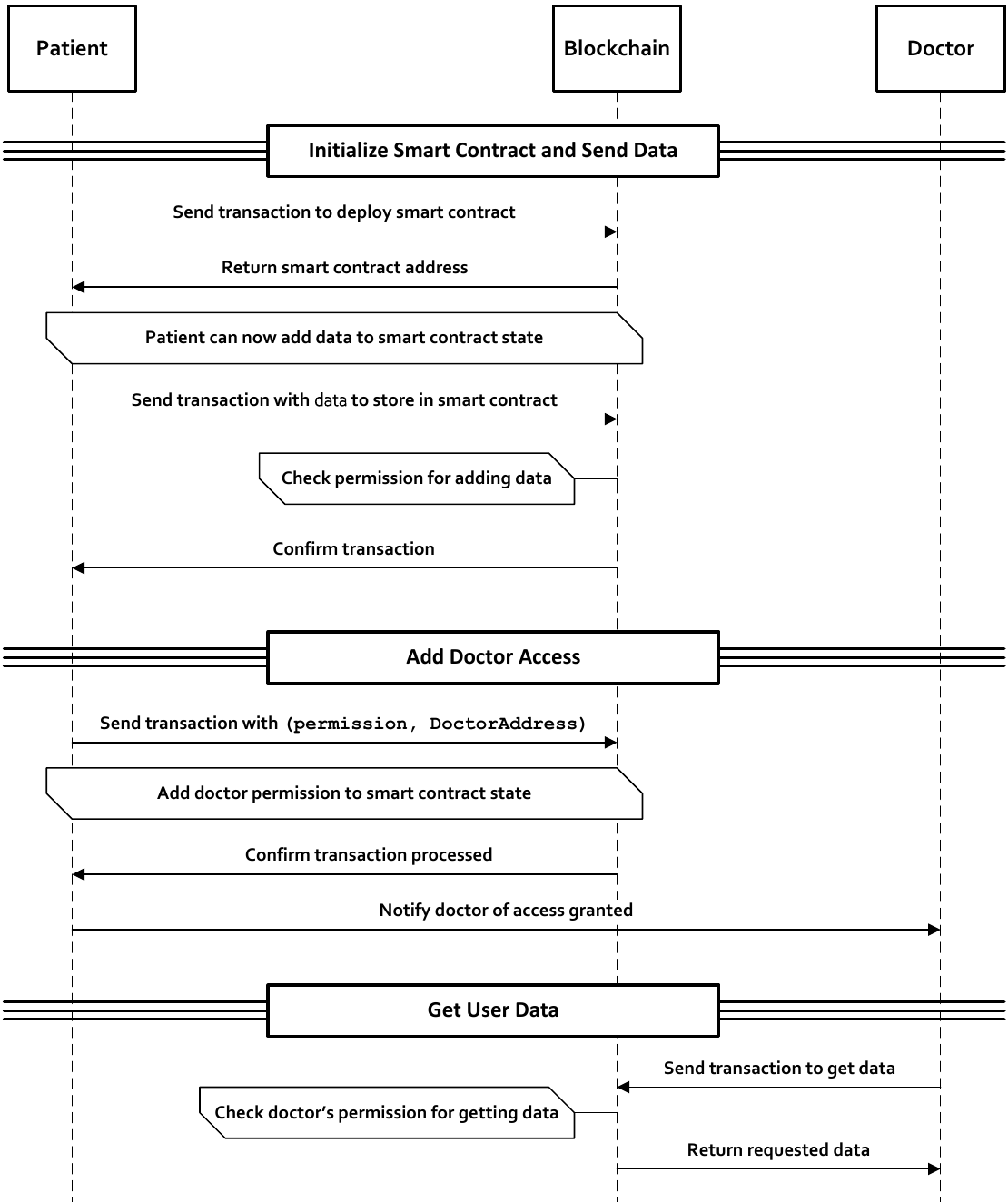}
			\caption{Sequence of Requests in the Experimental Scenario.}
			\label{figure2}
      \vspace{-13pt}
		\end{figure}
		
		\begin{figure*}[t]
			\includegraphics[width=\linewidth]{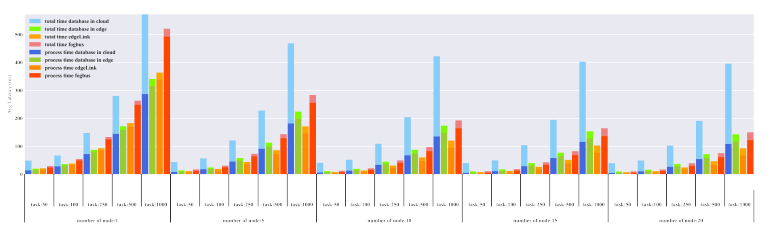}
			\caption{{\small Comparison of Processing Delay and Processing Time for Reading.}}
			\label{figure3}
   \vspace{-13pt}
		\end{figure*}
		
		\begin{figure*}[t]
			\includegraphics[width=\linewidth]{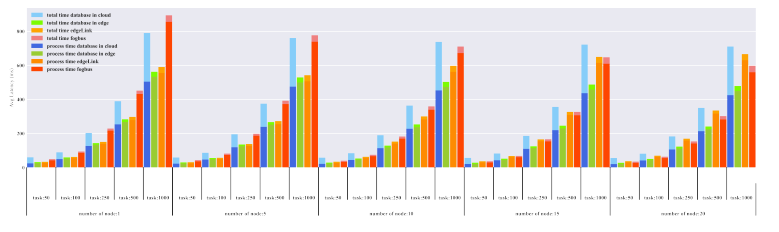}
			\caption{{\small Comparison of Processing Delay and Processing Time for Writing.}}
			\label{figure4}
      \vspace{-13pt}
		\end{figure*}
		
		\subsection{Security Analysis} \label{section.eval.sec}
			\noindent
			To examine potential threats in the EdgeLinker, we assume that attackers may be present at various layers, such as IoT devices, network nodes, or cloud storage. These attackers can eavesdrop on communications, create fake blocks and transactions, alter or delete data in storage, analyze multiple transactions to identify a node, and sign fake transactions to validate nodes. However, without possessing the private key, they cannot decrypt the encrypted messages.
			
			\subsubsection{Response of the EdgeLinker to Security Threats} \label{section.eval.sec.response}
			\noindent
			\textbf{Confidentiality:} In the EdgeLinker, user information is stored in a smart contract with access control, and accessing this information requires a pre-defined access unit in the smart contract. Additionally, if an attacker intercepts the transmitted messages, all messages are transferred through a secure channel that cannot be read by the attacker. Therefore, from the time of data generation to consumption, the attacker cannot access the confidential content.
			
			\textbf{Data Integrity:} The use of a hash, which is sensitive to the slightest change in all generated messages and blocks, reveals any tampering or alteration.
			
			\textbf{Availability:} Due to the distributed nature of blockchain, if an issue arises with one node, it can utilize other nodes for communication and information retrieval.
			
			\textbf{Authentication:} Each device has a private and public key. When sending a message through a secure channel, the sender signs the message hash with their private key, and the receiver can verify the authenticity of the sender by checking this signed hash.
			
			\textbf{Non-repudiation:} All information exchanges are conducted through a secure channel. Therefore, each exchanged message contains a signed hash that identifies the sender of the message. Additionally, during the block creation phase, the miner generates a signed hash of the entire block, which identifies the processor of the message.
			
			\textbf{Denial of Service:} This mechanism quickly drains the account balance of the attacking device, and then transactions sent from that user are no longer processed. Therefore, the attack is halted after a short period.
			
			\textbf{Eavesdropping:} All messages pass through a secure communication channel that encrypts the content of the message, preventing the content from being read.
			Replay Attack: In the secure communication channel, messages include a nonce and a timestamp, which help in detecting duplicate requests.
			Packet Dropping Attack: Currently, there is no solution for this attack. However, in the future, an alert system can be used to monitor communication intervals. If a communication disruption is detected, the message sending algorithm can be modified, and an alert can be sent for further investigation.
			
			\textbf{Identity Spoofing:} Since all communications are conducted through a secure channel, identity spoofing requires the individual's private key, which is not easily accessible.
			
			\textbf{Insertion Attack:} In this attack, the attacker gains access to the fog layer and creates blocks with incorrect transactions to form a false chain. However, since other nodes become aware of the fraudulent transactions during the new block validation phase, they reject the block and prevent it from being added to their ledger. Thus, this attack also fails.
			
			\textbf{Linkage Attack:} In this attack, attackers link blockchain transactions to an identifier to reveal the true identity of an anonymous node. To mitigate this threat, devices in the system can periodically change their keys, making it more difficult to associate transactions with a real identity. However, this does not completely thwart the attack.
			
		\begin{figure}
			\centering
			
			\begin{subfigure}[t]{0.23\textwidth}
				\centering
				\includegraphics[width=\textwidth]{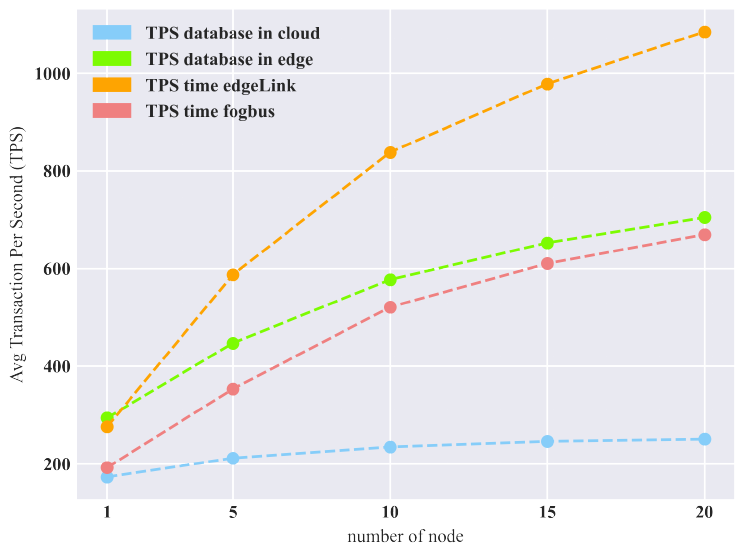}
				\caption{{\footnotesize Read Throughput Analysis.}}
				\label{figure5.1}
			\end{subfigure}%
			~ 
			\begin{subfigure}[t]{0.23\textwidth}
				\centering
				\includegraphics[width=\textwidth]{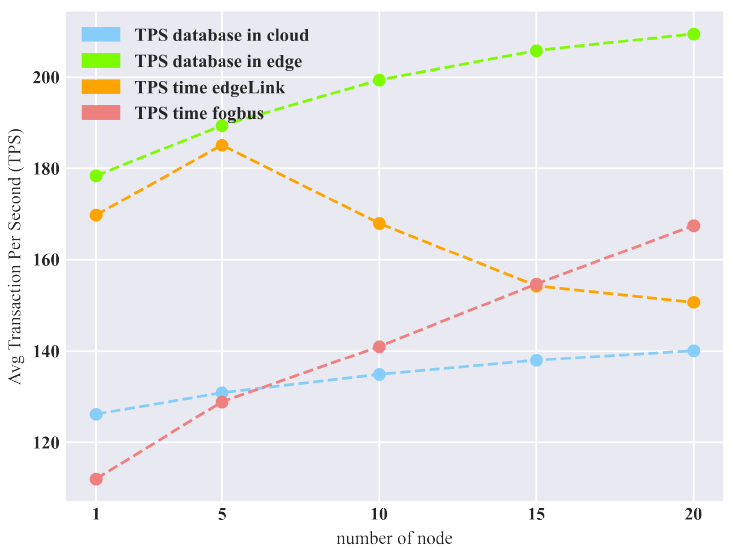}
				\caption{{\footnotesize Write Throughput Analysis.}}
				\label{figure5.2}
			\end{subfigure}
			
			\begin{subfigure}[t]{0.23\textwidth}
				\centering
				\includegraphics[width=\textwidth]{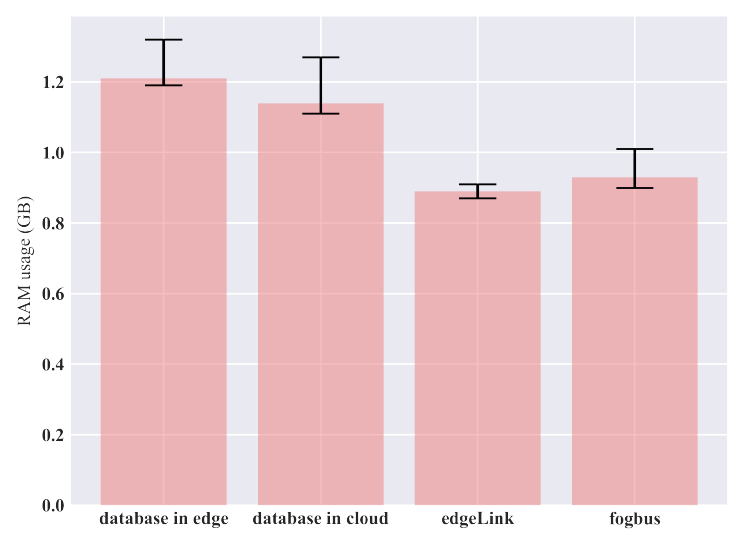}
				\caption{{\footnotesize RAM Consumption Analysis.}}
				\label{figure5.3}
			\end{subfigure}
			~ 
			\begin{subfigure}[t]{0.23\textwidth}
				\centering
				\includegraphics[width=\textwidth]{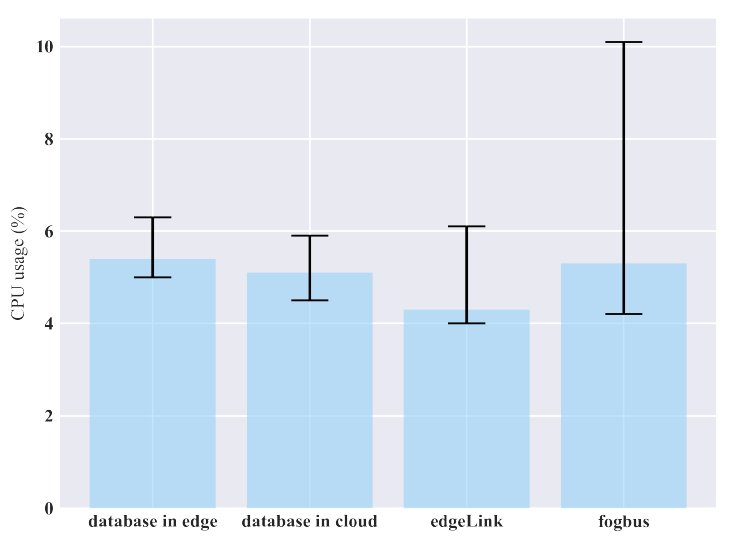}
				\caption{{\footnotesize CPU Consumption Analysis.}}
				\label{figure5.4}
			\end{subfigure}
			~ 
			\begin{subfigure}[t]{0.23\textwidth}
				\centering
				\includegraphics[width=\textwidth]{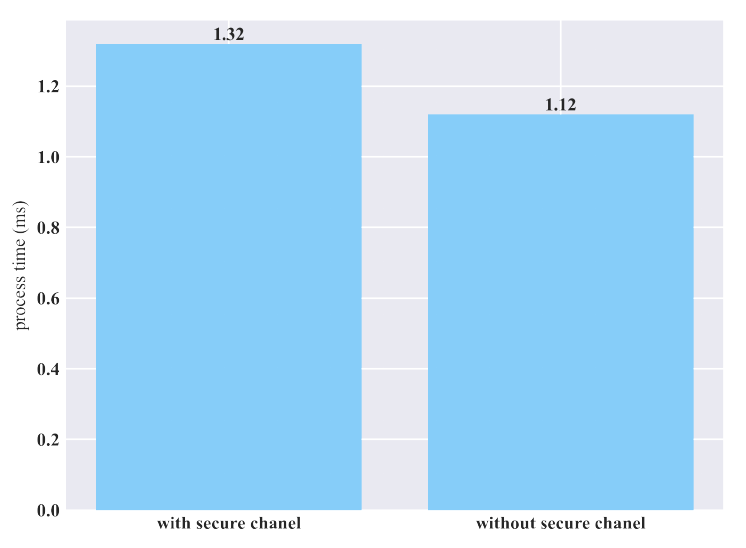}
				\caption{{\footnotesize Comparing Message Transmission Time Overhead in Baseline (right) and Secure Communication Channel (left) with EdgeLinker.}}
				\label{figure5.5}
			\end{subfigure}
			~ 
			\begin{subfigure}[t]{0.23\textwidth}
				\centering
				\includegraphics[width=\textwidth]{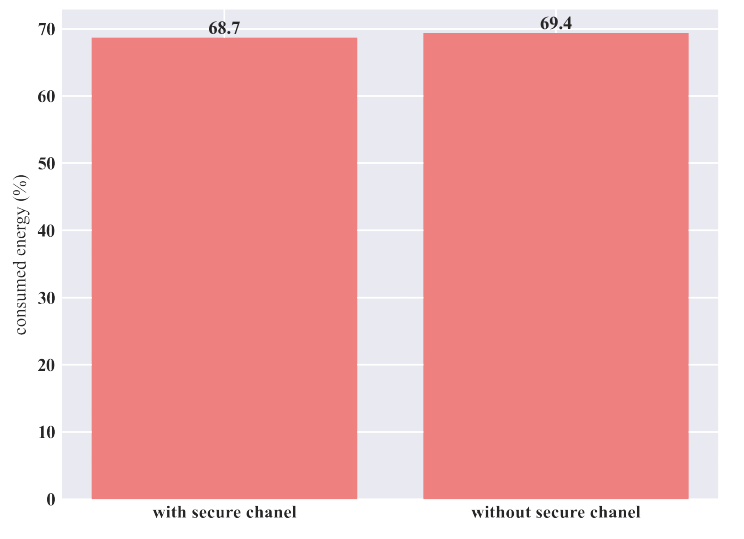}
				\caption{{\footnotesize Comparing Energy Consumption in Baseline (right) and Secure Communication Channel (left) with EdgeLinker.}}
				\label{figure5.6}
			\end{subfigure}
			
			\caption{{\small Evaluating EdgeLinker from Various Aspects: This evaluation includes throughput (\ref{figure5.1} and \ref{figure5.2}), CPU and RAM consumption (\ref{figure5.3} and \ref{figure5.4}), time overhead (\ref{figure5.5}), and energy consumption (\ref{figure5.6}).}}
			\label{figure5}
			
		\end{figure}

		\subsection{Performance Analysis} \label{section.eval.perf}
			\noindent
			The performance of the EdgeLinker was compared with a centralized cloud database, a distributed fog database, and FogBus\cite{b19}. The results indicate the superiority of the proposed solution in terms of latency and scalability. For greater accuracy and reproducibility of results, the statistics presented are the averages of 5 runs.
			
			To evaluate the performance of the EdgeLinker, three key metrics were considered: processing delay, processing time, and throughput (TPS). Processing delay is defined as the time required to record and store or retrieve transaction information in the blockchain or storage location from the moment the data is received by the node until the processing is complete. Processing time is considered as the time interval between the user's request and the receipt of the response. Throughput refers to the number of transactions that can be validated, executed, finalized, and confirmed per second after reaching consensus in the network. Measuring these metrics is essential for comparing the efficiency of different methods, and we also intend to compare the EdgeLinker in terms of energy consumption with other methods.
		
			Based on the results from the experiments shown in Figure \ref{figure3}, the read processing time with 5 nodes performs on average $23.8\%$ better, with 10 nodes $31\%$ better, with 15 nodes $33.2\%$ better, and with 20 nodes $35.1\%$ better compared to the edge database. Moreover, Based on the results shown in Figure \ref{figure5.1}, in all four methods examined, the throughput improves with an increase in the number of nodes. However, in the three methods—fog database, cloud database, and FogBus—due to the presence of bottlenecks, this increase occurs at a slower rate, and after reaching the maximum capacity of the nodes, the rate of increase gradually decreases.
			
			Based on the results shown in Figure \ref{figure4}, with 1 node, the fog database performs on average $4.7\%$ better than the EdgeLinker; however, with the increase to 5 nodes, this difference decreases to $2.3\%$, and the EdgeLinker performs similarly to the fog database. As the number of nodes increases to 10 or more, due to the need for more data exchange to achieve a distributed chain and finalize the created block, the capacity of the EdgeLinker decreases. In the FogBus solution, with the increase in the number of tasks, due to the PoW consensus algorithm, the average time to reach the block creation challenge response becomes closer to its average. Therefore, with 15 nodes and more than 500 tasks, it performs better than the EdgeLinker, but it should be noted that this difference is less than $0.2\%$.
		
			In Figure \ref{figure5.2}, the throughput of the EdgeLinker increases with the number of nodes up to 5 nodes, but thereafter, due to the increased coordination cost among the existing nodes, the throughput of the EdgeLinker decreases. In contrast, the FogBus solution, due to better load distribution among the incoming nodes, can increase its capacity and throughput at a higher rate compared to other solutions as the number of nodes increases.
			
			Regarding RAM consumption, as shown in Figure \ref{figure5.3}, the fog and cloud database solutions consumed more resources compared to other solutions. This is because databases, by default, use available memory capacity as cache and include certain background tasks required for data maintenance. The FogBus solution also requires more memory than the EdgeLinker due to the need for additional computations to solve puzzles. Among the various solutions, the proposed solution has the lowest RAM requirement, using these resources to cache newer blocks and maintain the overall state of the blockchain to improve speed.
			
			Additionally, based on the results shown in Figure \ref{figure5.4}, the FogBus solution consumes more CPU compared to our work when performing the heavy computations required to create a new block.
			
			Due to the use of encryption and hashing mechanisms in the secure communication channel, more time is required to process and send larger packets compared to the baseline. As shown in Figure \ref{figure5.5}, this increase in time is only about 0.2ms, which is negligible compared to the total processing time.
			
			The energy consumption of devices includes the processor, network, sensors, and display. The secure channel technique increases the packet size and requires more computations. This results in an average increase of 0.7 in energy consumption for transmission compared to the baseline method. To measure this, the device was initially charged to 80\% and 60 messages were sent at 1-minute intervals. The remaining charge was then measured. Figure \ref{figure5.6} shows these values.
			
			If we intend to implement EdgeLinker on an existing public network such as VeChain, each blockchain transaction on VeChain incurs a cost known as gas. This cost is determined based on the complexity of the smart contract operations and the network's congestion. The payment for gas is made using the VTHO token, where 1000 gas equals 1 VTHO. To measure the costs, we first determine them in a local test environment. Then, considering the price of \$0.001 per VTHO, we calculate the approximate cost of implementation on the real network. Table \ref{table2} provides a summary of the deployment and interaction costs with the smart contract.
			
			\begin{table}[t]
				\caption{{\small Cost of Deployment and Use of Smart Contract.}}
				\label{table2}
                \renewcommand{\familydefault}{\sfdefault} 
			\fontfamily{phv}\selectfont 
				\centering
					\begin{tabular}{|c|c|c|}
						\hline
						\textbf{Cost in USD} & \textbf{Gas Cost} & \textbf{Operation}\\
						\hline
						0.71 & 701382 & Smart Contract Deployment\\
						\hline
						0.05 & 48182 & Adding New Data\\
						\hline
						0.02 & 23521 & Granting Permission\\
						\hline
						0.02 & 21948 & Revoking Permission\\
						\hline
				\end{tabular}
     \fontfamily{\rmdefault}\selectfont 
        \vspace{-10pt}

			\end{table}

	\section{Conclusion} \label{section.conclusion}
		\noindent
		Given the nature of healthcare data, which is primarily studied by researchers and doctors with fewer new data entries, the proposed framework can be considered a practical solution due to its excellent performance in data reading. Additionally, considering the lower cost and system load of this framework, its capability to support legacy devices, and the ability to support user-customized applications, it can improve existing processes and facilitate the operational deployment of blockchain in the healthcare sector.

\end{document}